# Approaches to Modeling Insurgency

Alexander Kott, Bruce Skarin

## Introduction

More writings about insurgency appeared in the last few years than in the preceding hundred years (Kilcullen 2008). The explosion of interest in the subject has much to do with international interventions: insurgency is the single most difficult and frightening challenge that an intervention—military or non-military—may face.

For the purposes of this paper, we define insurgency as an organized movement that uses armed violence to overthrow a country's government while often hiding within the civilian population and using civilians to perform combat support functions. The use of civilian population differentiates insurgency from the regular warfare where such an exploitation of civilians would constitute a war crime. Similarly, a rebellion where anti-government forces do not disguise themselves as civilians and fight as a regular, identifiable military is different from insurgency. The significant involvement of civilian population also distinguishes insurgency from a purely terrorist movement, which relies primarily on a tight network of professional terrorists. Although our definition, like any other (e.g., US DoD, 2007), leaves room for gray areas, it serves to emphasize the key feature of insurgency—its reliance upon, and exploitation of civilian population. Because the literature on modeling, simulation, and analysis of regular warfare is vast and readily available, and because insurgencies are often associated with interventions, in this paper we limit our discussion to insurgencies.

Insurgency forces may include a combination of the following:
- an ideology-based movement that fights to overthrow the current form of the country's government and to establish a different regime;
- a personality-based movement driven to install its leader as the ruler of the country;
- a religious movement that wishes to defend its religious freedoms or to establish a religion-based regime in the country;
- an ethnic minority demanding greater rights or independence;
- a regional movement demanding secession or a greater share of the country's resources;
- an ethnic majority fighting against the rule of an ethnic minority or a colonial power.

Counterinsurgency forces also take a variety of forms:
- a democratic state that enjoys the support of a majority of the population;
- a dictatorship that relies on coercion to maintain its rule;
- a colonial government that represents a foreign power;
- a state that receives a limited support of a foreign power but is independent in its actions and could conceivably survive on its own;
- a state largely reliant on resources and support of a foreign power.

Regardless of the forces on the insurgent and counterinsurgent sides, importance and effectiveness of insurgencies have grown since the Second World War, for numerous reasons. These include the reluctance of Western or Western-supported governments to apply the brutal methods common in prior centuries; the effectiveness, low cost and ease of use of modern small arms like the Kalashnikov rifle (Singer ,2006); the easy availability of arms from a range of state and non-state supporters through channels of modern commerce (Anderson, 2007).

An international intervention can be a response to an insurgency, either in support of the insurgent side, e.g., African Union peacekeeping in Darfur since 2004 (Lynch, 2007), or in support of the counterinsurgents (e.g., the US support to the Colombian government fighting the FARC insurgents [Marcella, 2003]). On the other hand, an international intervention is often a cause of an insurgency, or a major factor in changing the insurgency's intensity or character. Thus, a change in insurgency can be both a cause and an effect of an intervention.

For example, a diplomatic intervention may induce a third party to discontinue its support to an insurgency; or compel a counterinsurgency-fighting government to conciliate with insurgents. An international famine aid or economic development assistance may reduce populations' grievances and its support to insurgents; but it may also increase the resources available to insurgents through protection racket (Baker, 2009).

Similarly, an international informational campaign that condemns an oppressive government may fan the flames of insurgency against the government; yet a campaign in support of a government may convince a part of the population that the government is an illegitimate foreign puppet. Finally, a military or law-enforcement intervention is likely to cause popular resentment at foreign meddling, or drive a segment of population to insurgency by depriving them of their prior privileges and wealth.

In turn, insurgency affects other phenomena (Kott and Citrenbaum, 2010). Economic development suffers, and illicit economy flourishes. Political dynamics shifts toward the competing positions on the issue of how to fight or to accommodate the insurgency.  Information channels become key tools—and casualties—of insurgents and counterinsurgents. Crime and corruption multiply as all sides may resort to bribes, death threats, protection racket, drug revenue, ransom and extortions. Ethnic, social and religious divisions are exploited and magnified in an insurgency.

## Qualitative Theories and Models

Theorists and practitioners of insurgency and counterinsurgency have outlined a number of key factors that affect the strengths of insurgency. Lenin (Osanka, 1962) stressed the importance of economic and social discontent of masses as a precondition to successful insurgency, as well as presence of a well-organized core of revolutionaries able to mobilize and guide the insurgency. Lawrence (1920, 1935) emphasized the need for an insurgent base inaccessible to the counterinsurgents forces, with protective terrain, adequate supplies of munitions, and at least passively supportive population—a safe haven where insurgents can hide and regroup. He also noted that insurgents benefit when counterinsurgent force relies on a vulnerable technology, such as a railroad.

Galula (1964) wrote about the critical role of civilian population who tends to be largely neutral in the conflict and shifts its support to insurgents or counterinsurgents depending on perceived benefits and outcomes of such a support. The population's support also depends on the actions, such as assistance or violent reprisals, taken by either side toward the population. Malayan insurgency (Nagl, 2002) offered the evidence that insurgency loses its strengths when population is physically separated and protected against the insurgents; when counterinsurgents offer economic benefits, security from violence, and political conciliation to the population. In addition, counterinsurgents benefit when they are able to attract a large fraction of population by exploiting ethnic and other differences. Indigenous counterinsurgent forces are more effective than foreign counterinsurgency forces in gaining population's loyalty.

Leites and Wolf (1970) point out that insurgency declines when deprived of resource inflows (such as munitions, supplies, and finances) and when its organizational structure and competency are disrupted by counterinsurgents. Respect and fear of government and its forces are important to dissuade population from supporting insurgents (Peters, 2006). War-weariness and anti-war sentiments among the counterinsurgent population and government may encourage and strengthen the insurgency (Iyengar and Monten, 2008; Anderson, 2007). Amnesty, financial rewards and offers of government and military positions can induce insurgents to switch sides (Kahl, 2007).

While the aforementioned factors are the most common drivers of insurgency, many other phenomena are important in specific situations. For example, a large pool of displaced persons or refugees can become a highly productive recruiting ground for insurgents as well as an opportunity to skim the foreign food aid (Cuny and Hill, 1999). Large-scale international economic aid programs can become the primary financing mechanism for an insurgency, through protection racket (Baker, 2009).

In an attempt to integrate a range of theoretical findings and practical observations, the US military produced a counterinsurgency manual (US Army, 2006), which is in part a comprehensive qualitative model of insurgency. The widely-cited manual identifies multiple factors that encourage and discourage insurgency, stresses that application of force can be a major factor in increasing population's resentment of counterinsurgency, and highlights population's security, good governance and essential services as key factors that diminish the population support to insurgency.

Unfortunately, empirical support to qualitative theories of insurgency tends to be anecdotal rather than scientifically rigorous. The work by Iyengar and Monten (2008) is a relatively uncommon example of a model-based, quantitative examination of a qualitative theory. These authors test the argument that anti-war sentiments in the USA emboldens the anti-US insurgents in Iraq and influences them to increase the rate of attacks on the US forces. Iyengar and Monten construct a theoretical model that relates the behavior of Iraq insurgents, specifically the rate of attacks on US forces and on Iraqi Government forces, to their perception of anti-war sentiments in the US. In this model, insurgents are rational, strategic actors who attempt to optimize the distribution of their attacks over time in such a manner that the insurgents preserve their resources while maximizing the anti-war opinions in the US. The authors compute the differences in predictions of the model for different areas of Iraq—some with greater access to information about the US public opinions than others—and compare these estimates with the reported insurgent attacks.  They find that in periods immediately after the US media reports a spike in anti-war sentiments, the level of insurgent attacks increases.

Also unfortunately, interpretation and application of qualitative models to practical decision-making is an imprecise art. When in 2007-2007 the US decision-makers pondered whether to increase or to decrease the number of US troops of Iraq—the so-called Surge decision (Woodward, 2008)—the qualitative theory was hardly in question. Most likely, all participants in the debate agreed that increasing the number of US troops fighting the Iraqi insurgency may improve the security for a fraction of the Iraqi population; it may also increase the population's anger at foreign occupation; it may also give the Iraqi government additional time to strengthen its political and military posture; or, it may also lull the government into complacent reliance on the US protection. However, the decision-makers and consultants disagreed strongly on the relative quantitative magnitudes of these potential qualitative effects, and on the resulting balance.

The overwhelming majority of the US senior military leaders believed that on the balance the Surge—rapid temporary injection of additional US troops into the counterinsurgency efforts—would be counter-productive because it would merely encourage the Iraqi government to continue its complacent dependency on the US (Woodward, 2008; pp.224-281). A small group of civilian theoreticians and retired generals believed otherwise and urged President George W. Bush to accept the Surge plan.

In the event, President Bush decided to execute the Surge, and a major reduction of insurgency followed a few months later, in the middle of 2008. Opinions still differ on whether the Surge worked as its proponents expected, or whether other, unrelated mechanisms caused the reduction in insurgency (Woodward 2008; pp.380-384; Pierson 2008).  Qualitative models are insufficient to answer such questions; they require quantitative models with the corresponding quantitative metrics, variables, and relationships.

## Quantitative Measures of Insurgency

To construct quantitative model of a complex phenomenon, such as insurgency, one needs ways to measure attributes and dynamics pertaining to that phenomenon. Formulating meaningful metrics of insurgency, however, is a significant challenge. Insurgencies are largely about human perceptions, which are contextual. For example, public opinion about the quality of current situation in a country is highly dependent on past historical experiences and availability of alternatives. Thus, interpretation of a metric's magnitude or event its trend is dependent on other, often intangible variables (Campbell et al, 2009).

Most commonly used metrics of insurgency measure level of violence, e.g., the number of insurgent attacks per month; quality of government institutions, e.g., public opinion polls regarding the level of corruption; and strengths of security forces, e.g., the number of counterinsurgent troops and their degree of readiness. For example, Brookings Institution offers comprehensive datasets of metrics (O'Hanlon and Campbell; Campbell and Shapiro) for insurgencies in Iraq (since March 2003) and Afghansistan (since October 2001). These datasets include several dozens of metrics such as fatalities and counterinsurgent troops, number of insurgent attacks of different types, strength of counterinsurgent troops, strengths of anti-insurgent militia, unemployment, electricity generation, inflation, GDP, and public opinion polls.

Others begin to explore more comprehensive processes of measuring insurgency, with a special focus on the insurgency's less-tangible aspects like population attitudes and perceptions. For example, the MPICE program (Dziedzic, 2008) has developed a broad-ranging recommendation for gathering a variety of in-depth metrics with computer tools. These would include semi-automated analysis of a country's media content to gauge popular and elite impressions of insurgency-related issues; creation of a panel of experts to assess issues of interest (e.g. the capacity of law enforcement agencies to perform essential functions); and specially constructed public opinion surveys.

There is no shortage of complaints about metrics being potentially meaningless and even misleading. For example, Clancy and Crosett (2007) describe history of several insurgencies and find that metrics used in those insurgencies were highly misleading. Still, critics of metrics agree that analysts and decision-makers must look for insightful metrics, and for better means to interpret their meaning. Quantitative models can help do exactly that.

## Influence Diagrams

Also called the causal loop diagram, influence diagram occupies the middle ground between a qualitative model and a quantitative model. Like a qualitative model, an influence diagram describes key aspects of insurgency phenomena and the influences between them. In addition, however, an influence diagram offers features that make it a stepping-stone toward a quantitative model: the diagram names specific quantitative variables, identifies dependent variables for each variable, and specifies whether an increase in a variable causes an increase or decrease in its dependent variable.

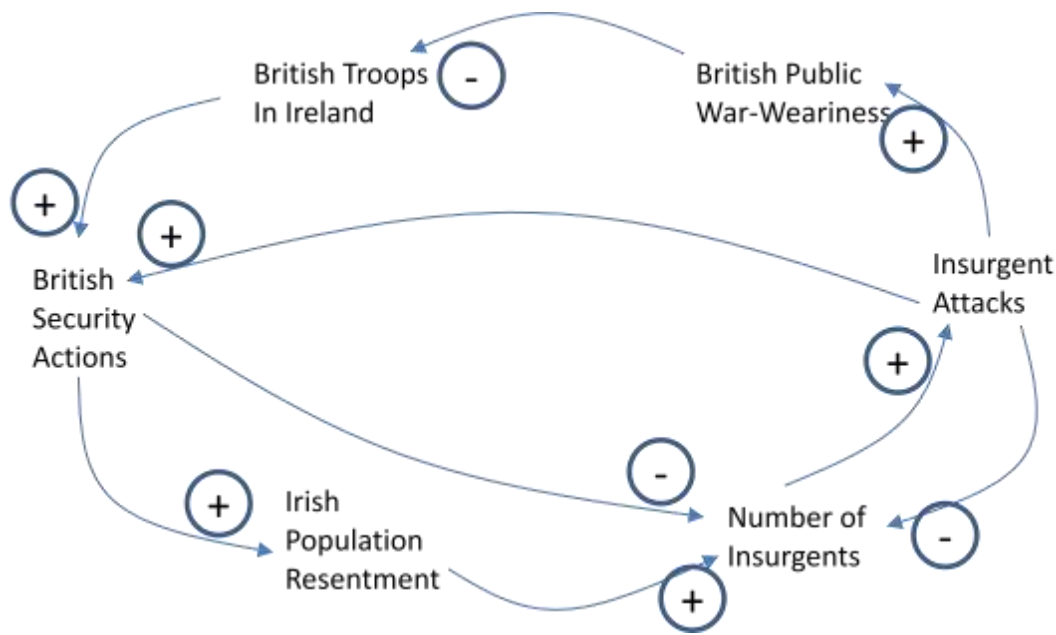

**Figure 1. Key variables and their relations for the insurgency of Anglo-Irish War 1916-1923**

In Figure 1, the influence diagram shows key variables and their relations that describe the insurgency of Anglo-Irish War (also known as the Irish War of Independence) of 1916-1923 (Anderson, 2007). Let us begin with the variable called Number of Insurgents, which reflects the number of active anti-British insurgents operating in Ireland. An increase in Number of Insurgents leads to increase in Insurgent Attacks—note that the two variables are connected with an arrow and marked with the plus sign (increase leads to increase).

The growing Number of Attacks in turn leads to an increase in British Public War-Weariness (again, the plus sign indicates that increase leads to an increase) and prompts the British forces in Ireland to energize their British Security Measures, which leads to greater Irish Population Resentment and also to more arrests that deplete the Number of Insurgents. Note the last arrow (from British Security Measures to Number of insurgents) is marked with a minus sign, because here increase leads to decrease. And so on.

The diagram points to potentially very complex non-linear dynamics of insurgency. Even in this simple model, the number of Insurgent Attacks, for example, affects the Number of Insurgents in five different ways—there are five distinct paths from Insurgent Attacks to Number of Insurgents.
The benefits of constructing such a diagram include the following:
- The modeler or analyst elucidates and formalizes her thinking about the insurgency phenomena;
- Specific variables and their relations are identified and documented;
- Qualitative nature (e.g., increase in A leads to decrease in B) of the variable dependencies are determined and documented;
- Complex feedback loops and side effects become clearly visible;
- Subject matter experts and other analysts and modelers can review and confirm or question the visual representation of the model.

Ideally, the modeler derives the influence diagram directly from relevant qualitative theories. For example, Pierson faithfully followed a single qualitative model--the Counterinsurgency Manual (US Army, 2006)--to build an influence diagram of Iraq insurgency (started in 2003), with a large number of variables and influence lines (Pierson, 2008).

Choucri et al (2006) formulate their influence diagram of insurgency while rigorously documenting social science theoretical literature in support of each of their model's influences. For example, instead of merely asserting as self-evident the influence "More Insurgents Lead to More Regime Opponents," they cite literature that supports existence of such an influence.

They also attempt to justify the validity of variables they introduce into their model. For example, they introduce a variable called State Resiliency and justify it by comparing the State Resiliency to the determinants of civil war of Hegre et al (2001).

## System Dynamics Models of Insurgency

The modeler may continue to develop the model of Fig. 1 by specifying equations that relate each variable to the variables that influence it, e.g., the equation that computes the Number of Insurgents as a function of British Security Measures and of Irish Population Resentment. The resulting system of equations (typically a system of coupled non-linear differential equations) can be solved, for example, by numerical simulation. The solution will shows how each variable evolves over time.

 System Dynamics (Sterman, 2001) is a technique that simplifies specifying and solving such systems of equations. A variable is represented as a "stock" of goods. Inflows and outflows represent temporal changes to the variable. A "valve" that opens and closes as a function of other variables controls the rate of a flow.

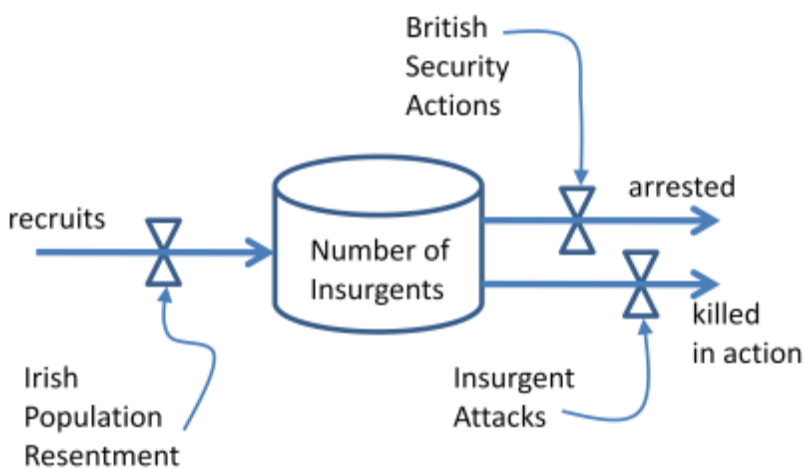

Figure 2. A fragment of a System Dynamics model

Fig. 2 depicts a fragment of a System Dynamics model that elaborates the influence diagram of Fig. 1. Here, the Number of Insurgents is a stock, or a level of liquid in a reservoir. The incoming pipe carries the flow of new recruits; the valve opens wider when the Irish Population Resentment is greater. One outgoing pipe represents the depletion of Number of Insurgents due to arrests. The valve opening on that pipe depends on the British Security Measures. The second outgoing pipe represents the number of insurgents lost in action, and the valve is controlled by the number of Insurgent Attacks. The modeler must specify an equation for each valve. A computerized System Dynamic tool such as (isee systems, 2009) helps to specify the model and then solves it automatically.

System Dynamics is arguably the most popular technique of insurgency modeling. For example, Fig. 1 is partially adapted from Anderson (2007) who constructed a system dynamics model of the Anglo-Irish War, possibly the first modern urban insurgency. Anderson used only a few causal loops—closed paths through a set of variables. One loop represents the insurgency suppression and creation: coercive acts of British forces increase interference in civil life, which increases population resentment, which increases number of insurgents and their anti-British attacks, which leads to increase in British coercive acts. Another critical loop reflects the impact of British war weariness: as insurgent violence increases, British public's war-weariness increases, leading to the public pressure to remove British troops from Ireland.

If Anderson (2007) exemplifies a high-level model capturing overall dynamics of the entire insurgency, the work of Grynkewich and Reifel (2006) is an example of a detailed model of a particular sub-feature of insurgency. They model the financial operations and organizational behavior of the Salafist Group for Preaching and Combat (known by its French initials, GSPC). The model relates intensity of insurgent combat operations, expenses to support the operations, and influence of combat operations on population's willingness to support insurgents financially.

The key stock in this model is the pool of finances available to GSPC; key inflows include extortion from population, voluntary donations, smuggling operations, kidnappings and ransoms. Outflows include organizational overhead and operational costs. Authors use limited published data and educated guesses to derive values of equation parameters such as the fraction of population willing to donate to insurgents and the amount of donations. The authors report that the model's simulations agree qualitatively with the available information regarding the GSPC's operations and finances.

While few would claim that a system dynamics model of insurgency provides a reliable prediction of an insurgency's future evolution, there are other significant benefits in constructing and simulating such a model:
- The model helps analysts and decision-makers see un-anticipated side effects, particularly those due to feedback loops;
- The modelers and analysts document systematically a formal model that includes a rich, integrated set of factors, processes, and quantitative dependencies;
- The simulation of the model illustrates the complexity of the nonlinear temporal dynamics of the insurgency system;
- Sensitivity analysis aids analysts and decision-makers in forming insights and intuition toward formulation of intervention plans and policies.

# Agent Based Modeling of Insurgency

Let us return to Fig. 1. Even in this simple diagram, we see several distinct actors, each with its own set of goals, actions, culture, resources, and relations: insurgent organization, general Irish public, British forces in Ireland, British public and British government. We may conclude that this set of actors (agents) is too simplistic: after all, Irish insurgents included various movements with different tactics and leadership, the British public included pro-war and anti-war segments, and the British government included parties with different views on the Irish question, and so on.

If we wish to model the dynamic relations and mutual influences of all these agents, we should add multiple variables associated with each of the agents, and influence lines between the additional variables. The model becomes unwieldy.

An alternative is to use the agent-based modeling paradigm. An agent-based model consists of agents – software representations of individuals or groups of individuals. Groups can be represented at different scales of abstraction: organizations, segments of populations, ethnic or religious groups, social classes, political parties, or movements, even whole countries.

A model may also include elements of the agents' environment. For example, we may want to represent significant geographic areas where the insurgency unfolded: Dublin, Southern Ireland, Northern Ireland, etc.

An agent has attributes, e.g., an insurgent organization may be characterized by the number of members, amount of munitions, funds available, level of combat training, political objectives, and organizational competency. An agent has relations to other agents, e.g., an insurgent group may be an ally of another insurgent group.

An agent has means by which to make decisions about the actions it will take. Computationally, agents can be implemented, for example, as objects. In that case, an agent has methods by which it makes decision, i.e., choices among available actions. Such a method may include rules or decision-making algorithms, stochastic or deterministic.

An agent has a set of actions it can take, e.g., bombing the barracks of counterinsurgent forces, moving itself to another county, or adopting a more positive attitude toward a rival insurgent group. When executed, an action affects attributes and relations of this and other agents. For example, a bombing attack by an insurgent group reduces the strength and tolerance level of counterinsurgent agent, reduces the amount of explosives available to the insurgent group, and increases its morale and reputation for effectiveness.

To construct a complete model, the modeler identifies the appropriate set of agents (possibly starting with an influence diagram like Fig. 1), assigns attributes and relations for agents, and codes the methods for agent decision-making and for action impacts. Prior to executing the simulation of the model, the modeler also assigns initial values (i.e., the values at the start of the simulation) of the attributes and relations.

The simulation of the model usually proceeds in time-steps. A time-step for modeling insurgency is often a month or a week. At the beginning of the simulation, at week-1, each agent uses its decision-making method to select one or more actions. Then each agent executes the selected actions and each action modifies values of appropriate attributes. This completes the first time-step, and the process repeats for the next time-step, week-2, and so on. Because attributes and relations of agents change with time, in each time-step agents may select different actions (or no action).

The simulation ends when agents reach the last time-step. Usually, the analyst who uses the model specifies the number of time-steps. For example, if the analyst studies the potential insurgency effects of an intervention campaign that is to last five years, she may specify 60 time-steps, with each step corresponding to a month of time.

At the end of the simulation, the analyst reviews the history of the simulated agents: changes in their attributes and relations over time. For example, an analyst may observe that an agent representing an insurgent group rapidly increases its strength between months 1 and 15, then begins to lose support of local population between months 15 and 20, rapidly depletes its strengths between months 20 and 23, and finally merges with another insurgent group at month 27.

Depending on the tool or model used, an agent may have a memory; it may accumulate experiences, learn new rules, and change its beliefs. For example, the CORES system (Kowalchuck et al, 2004) models an agent's belief in its own actions. When an action does not succeed, the agent's belief in the worth of the action diminishes. Thus, a counterinsurgency agent may gradually come to conclude that harsh retributions are not effective.

Nexus (Duong, 2007) agent-modeling tool pays even greater attention to the cognitive nature of its agents. A Nexus agent has a degree of historical consciousness; it assigns and reassigns blame for past actions of agents, changes beliefs in trustworthiness of other agents, judges their ideology and looks for friendship with its enemy's enemy. Nexus played a major role in a large-scale, real-world study by a US government agency, in a situation that involved potential international intervention and insurgency.

Senturion's agents possess a complex decision-making mechanism that comprises a set of algorithms drawn from game theory, decision theory, spatial bargaining, and microeconomics. Together, they model how agents interact in a political process. This tool has produced multiple real-world predictions of insurgency-driven situations, such as those in Iraq, Palestinian Territories, and Darfur in 2004-2009 (Abdollahian, 2005; Sentia, 2008).

## Other Modeling Methods and Tools

Broad overviews of tools relevant to insurgency modeling are found in (Hartley, 2008; Benedict and Simmons, 2007). Virtually all tools able to generate an anticipatory estimate of an intervention's impact on insurgency fall into one of the two categories we already discussed: system dynamics modeling or agent-based modeling. However, one finds a few exceptions that fall into two other categories: human-driven wargaming and statistical correlations.

### Human-in-the-Loop Wargaming

PSOM model (Parkman, 2005) is an example of a wargaming-based approach—a computerized, time-stepped war-game where human players decide the actions and moves of insurgent and counterinsurgent forces. In PSOM, the geographic area of operations (the wargame board) is divided into 50 km squares. Each square has attributes such as its degree of urbanization, nature of terrain, population density, quality of infrastructure, cultural values, population's perception of security and support to the government.

Human players operate the insurgency and counterinsurgency forces. At the beginning of the wargame, the players allocate their respective force units to selected squares of the wargame board. Players assign particular missions to these force units: enforce, stabilize, disrupt, and others. During each time-step, the computer determines the outcome of each force unit's mission based on the current condition in the square, and on actions and strengths of the opponents' forces in the square. The outcome then leads to changes in the square's attributes values. For example, if the counterinsurgent force unit deems successful in its security-enhancement mission, the value of the security attribute in the square increases. Then the game proceeds to the next time-step, and so on.

One can notice here a certain similarity to the agent-based modeling, except that in PSOM the human players select the agents' actions (missions and moves), while in the agent-based paradigm agents make decisions without human intervention.

### Statistical Correlations

Application of statistical techniques to historical data on insurgencies yields valuable correlations. Some regularity in data is noticeable even without a formal analysis. For example, (Quinlivan, 1997) offers a compelling visual correlation between an intervention's success and the number of security personnel (military plus police) deployed per thousand of the country's inhabitants. Historically, successful suppression of an insurgency requires about ten or more security personnel per thousand of population.

Elbadawi and Sambanis (2000) offer a rigorous quantitative analysis of factors affecting the duration of civil wars (including insurgencies). They find that an external intervention tend to prolong a conflict. They also find a strong U-shaped correlation between the duration of an ethnically based conflict and the ethic fractionalization index. Conflict lasts longer in countries with two or few large ethnic groups than in those with many small groups or a single dominant group.

# Initialization, Calibration and Validation

An insurgency model requires initial values of variables and values of constant parameters or coefficients in equations or rules. The modeler also needs a validation process that shows an acceptable degree of agreement between the model's outputs and data or trends observed in the real insurgency.

Often, modelers have to make educated guesses based on very limited data. E.g., Grynkewich and Reifel (2006) are compelled to use a single newspaper quote of an unnamed "Hezbollah operative" to assign a cost to an insurgent operation.

When no insurgency-specific data are available, modelers resort to the use of data from domains partially similar to insurgency. For example, Robbins (2005) presents a system dynamics model for reconstruction and stabilization. The model includes an insurgency sub-module that accounts for such factors as ethnic fractionalization, effect of unemployment and urbanization. Lacking insurgency-related data, Robbins uses correlations obtained from studies of crime dynamics.

Others derive quantitative parameters using rather sophisticated models, extensive collection of real-world data and comprehensive statistical analysis. For example, Iyengar and Monten use such formidable arsenal of tools to quantify the degree of influence that the apparent lack of resolve among the US public has on the intensity of insurgent attacks.

In many cases, the modeler obtains a model's parameters by calibration, i.e., by changing the values until the model's outputs match the available data. For example, Leweling and Sieber (2006) calibrate their model of human resources of an insurgent organization against data derived from publically available news reports, such as numbers of insurgents arrested. They adjust both the structure and parameters of the model in order to obtain satisfactory agreement between the model's output and the data.

Although hardly the best practice, some modelers consider calibration identical with validation. The modeler calibrates the parameters of his model, shows a reasonable agreement between the model's output and the available real-world data, and then declares the model valid. Ideally, he should calibrate with one set of data and validate with respect to another set data. Often, unfortunately, only one set of data is available.

For example, Anderson (2007) validates his system dynamics model by comparing the model's results with the data describing the Anglo-Irish War of 1916-23. He uses this particular civil war for validation purposes because it is the first modern urban insurgency, and because it is a rare case of a well-documented insurgency. The model was able to replicate approximately the dynamic behavior of the Anglo-Irish War, suggesting a degree of model's validity. Anderson lists numerous parameters and values of these parameters without explaining how he obtained the values. One has to presume that he calibrated the values in a way that maximized the agreement between the model results and the real-world data.

## Case Study: Conflict in Northern Ireland, 1966-1998

The Republic of Ireland occupies about 80% of the island of Ireland. The remaining northeastern area of the island, the Northern Ireland, is a part of the United Kingdom. Between 1966 and 1997, the Northern Ireland experienced an armed conflict known as the Troubles (Tonge, 2006). Several irregular combatant groups rooted in Irish Catholic population, notably the Provisional Irish Republican Army (Provisional IRA) and the Official Irish Republican Army (Official IRA), fought for reunification of the Northern Ireland with the Republic of Ireland. Their opponents, irregular combatant groups of Protestant origins such as the Ulster Volunteer Force (UVF) and Ulster Defense Association (UDA), fought to maintain the Northern Ireland as a part of the United Kingdom. All irregular combatant groups were in conflict with the government of the United Kingdom and its armed forces. Because the irregular combatants disguised themselves as civilians and relied on widespread popular support, the conflict is best characterized as an insurgency.

Political parties in the Northern Ireland were strongly polarized along the lines of ethno-religious affiliation. Pro-British Protestant parties included the Ulster Unionist Party (UUP) and the Democratic Unionist Party (DUP). Supporters of these parties were more likely to sympathize with the UVF and UDA combatants. Irish Catholic population tended to support such parties as Social Democratic and Labor Party (SDLP) and Sinn Fein. The last one is often described as the political arm of IRA, and Sinn Fein's supporters were likely to assist IRA combatants (Silke, 1999).

Let us review in detail a model (Grier et al, 2008) that focuses on the Troubles in Northern Ireland starting in 1968. The model is agent-based and uses a modeling tool called Simulation of Cultural Identities for Prediction of Reactions (SCIPR). Our objective in this modeling effort is to predict trends in the degree of population's support to parties in this conflict. In effect, we ask the following question: if we were to have a model like this in 1969, could we predict trends in population's sympathies to political movements like DUP and Sinn Fein. Arguably, insurgents on both sides draw their strength from population segments that identify with extremes of the political spectrum. If we can predict trends in extreme political opinions, we are better prepared to anticipate changes in the strength of insurgency.

Another important question is how much data we need to construct and simulate such a model. Models that require less data are less expensive to construct, and easier to understand. In this case study, we use a rather simple model that requires little data. We find, encouragingly, that the simple model with a modest amount of readily available data produces potentially useful predictions of trends.

### Agents

Using SCIPR, we construct about 5000 agents that represent the entire population of Northern Ireland. Each agent represents a group of approximately 300 individuals with approximately similar identities, residing near the same locale.

An agent has several attributes including

- the district of residence (one of the 26 districts),
- ethno-religious affiliation (Catholic or Protestant),
- a number between 0.0 and 1.0 representing the agent's opinion on the issue of Northern Ireland affiliation (0.0 means Northern Ireland must remain British, 1.0 means Northern Ireland must unite with the Republic of Ireland), and
- the political party that the agent supports.

An agent has social links to an average of ten other agents, of which 90% are of the same religion. These networks are fixed and are formed under the assumption that individuals in Northern Ireland have around ten people that they are in regular contact with to discuss political issues, and that they are generally of a similar identity. Using these links, an agent can communicate its political opinion to other agents.

There is no particular theoretical basis for using 5,000 agents, and not 500 or 50,000, but for simulations of less than 1,000 agents we find the statistical variance of the generated agent population from the input distributions is significant. With simulations greater than 5,000 there is no significant difference in either the initialization or outcome. Therefore, the number 5,000 is merely a modeling assumption, and it seems to work for our purposes in this model.

Note what we do not attempt to model: we do not model political movements explicitly, nor political leaders, nor other influential countries like the Eire or the United States, nor the rest of the Great Britain. Neither do we represent explicitly the insurgency groups, counterinsurgency forces, economics, and many other potentially significant factors.

### Agent Actions

An agent can perform several actions:

- communicate its current political opinion to another agent through an existing link we mentioned earlier,
- change its political opinion on the question of Northern Ireland affiliation in response to receiving an opinion from another agent,
- change its political opinion on the question of Northern Ireland affiliation in response to the news of a latest sectarian killing, and
- change its party affiliation.

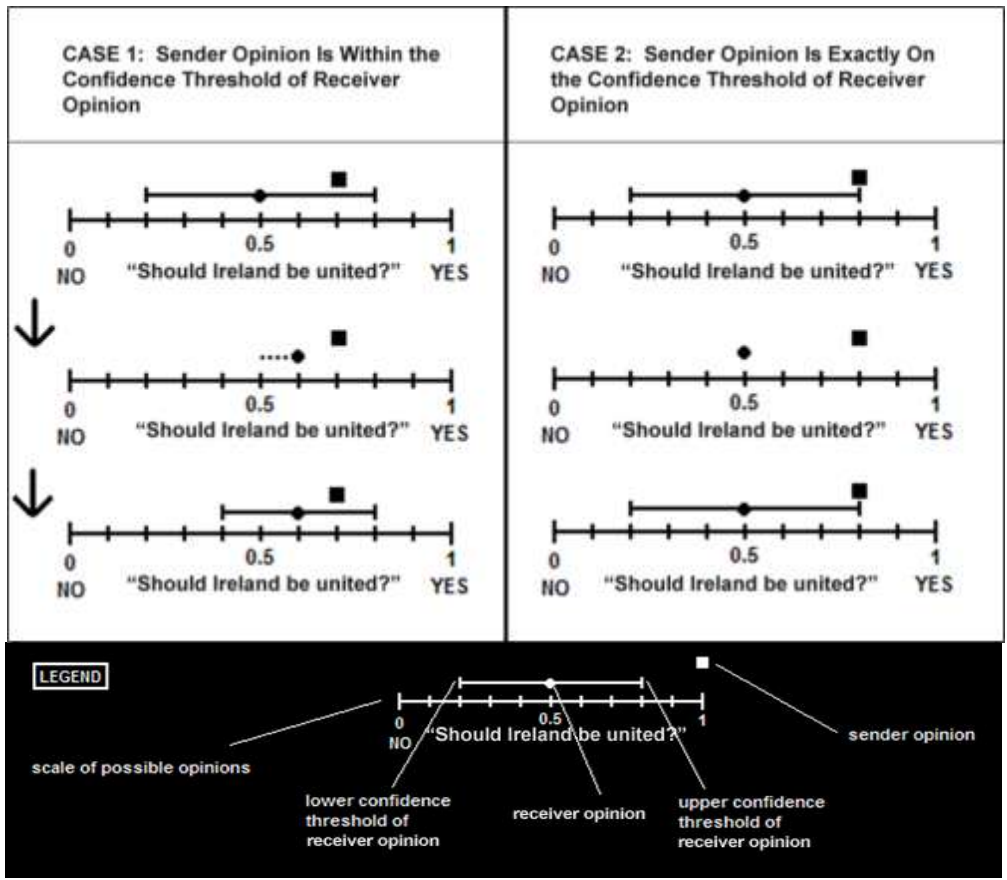

**Figure 3. A model of an agent's change of its political opinion.**

An agent changes its political opinion in the manner depicted in Fig. 3. An agent's opinion is characterized by the opinion number, e.g., 0.5, and the opinion's confidence bounds, e.g., (0.25; 0.75). When the agent receives an opinion, .e.g., 0.7, from another agent within his social network (i.e., connected by an existing link), the receiver modifies its opinion partially, in the direction of the sender's opinion. When the sender's opinion is outside the receiver's confidence bounds, the receiver ignores the sender's opinion. This model, and the opinion scaling equations, follows largely (Friedkin, 1999; Hegselman and Krause 2002).

An agent also changes its political opinion in response to events, in this case, the latest episode of sectarian killings. Responses to the event are specified by identities as a distribution of reactions to an opinion. After the occurrence of an event, agents within the region of the event sample a value from the distribution of reactions for their identities. This value is then used to scale the maximum opinion change parameter and added to an agent's current opinion. This new opinion is then evaluated using the same bounded confidence procedure described above. In this model, when for example a Catholic is killed, agents of the same religion increase the strength of their opinion in favor of the united Ireland.

An agent switches to support of another party when the agent finds its opinion more closely aligned with those of the members of another party than with the members of the agent's current party.

For the sake of clarity and brevity, we omit a few other relevant details, such as an agent's change of opinion in response to a sectarian murder.

## Model Initialization

We initialize our model to resemble the Northern Ireland of 1968 by assigning each agent the values of its attributes. We pick the values stochastically but our distributions are such that the total numbers of Protestant and Catholic agents, as well as fractions of supporters of each party in each county correspond to the demographic and voting data of 1968 (Fig. 4).

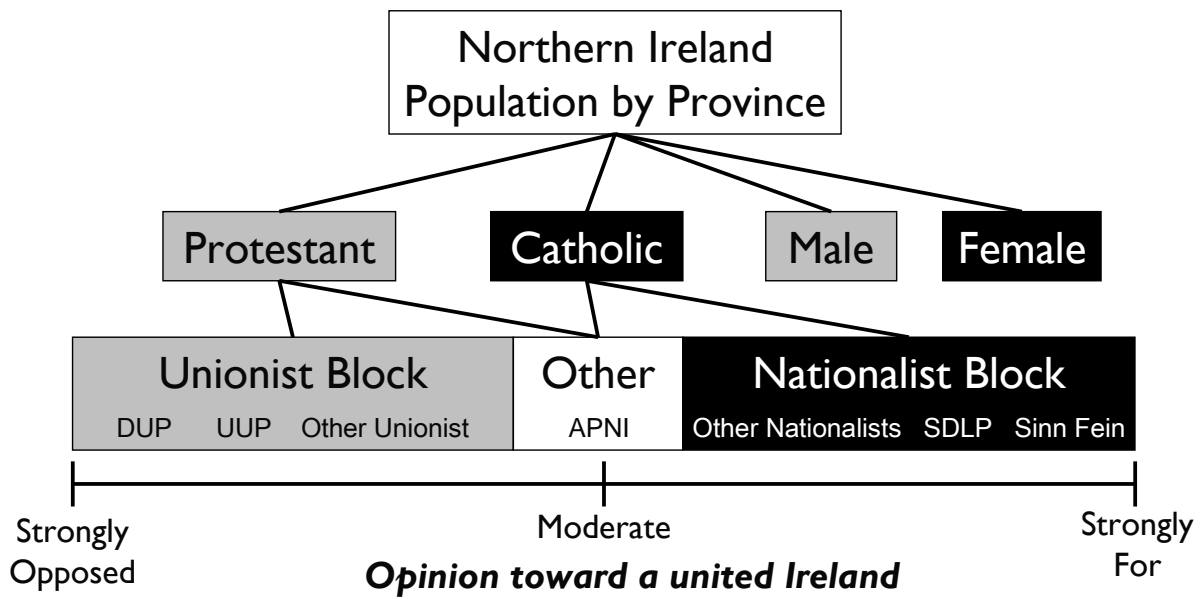

**Figure 4. Demographic and voting categories.**

We also create a social network of agents: we assign a link to a randomly chosen pair of agents in such a manner that each agent has on average about 10 links, with about 45% of the links connecting to agents within its immediate vicinity, 30% to other agents residing in the same district, and the rest---to agents in other locales of Northern Ireland. We also ensure that about 90% of the links are between agents of the same religion. These parameters—10, 45%, and 30%--are merely modeling assumptions, without a substantive empirical basis. A more rigorous modeling effort should consider obtaining empirical data to support and improve these assumptions.

We also initialize a table of sectarian killings: for each day, how many (if any) Catholics and Protestants perish in inter-communal violence. Our model does not predict such data; they have to come from a separate source.

Now we have the initial attributes of agents and their links fully defined; the model is ready for simulation.

## Simulation Process

The simulation algorithm begins its process on the first day of the year 1969, in simulated time. The algorithm randomly picks a number of pair of linked agents. In each pair, the algorithm randomly designates one of the agents to be the sender and another as the receiver of the political opinion. The number of pairs selected on each day is such that the every agent, on average, acts as a sender approximately every three days. This parameter—three days—is merely a modeling assumption; a more rigorous model should test the empirical validity of this assumption.

When selected by the algorithm, the sender communicates its current political opinion to the receiver. The receiver then either ignores the received opinion, or shifts its own opinion partially toward the sender's opinion as we discussed above.

The simulation algorithm also notifies each agent of sectarian killings that occurred on that day. Agents adjust their political opinions correspondingly.

The simulation algorithm then performs the same process on the next day of the simulated time, and so on. At the end of each year, each agent re-evaluates the alignment of its political opinion with the members of political parties. Depending on the alignment, the agent changes its party support. The algorithm stops at the end of the year 2005, as instructed by the modeler, and outputs the results.

## Results of the Simulation

Fig . 5 and 6 compare the historical results of elections and polls in Northern Ireland with the results of our simulation.

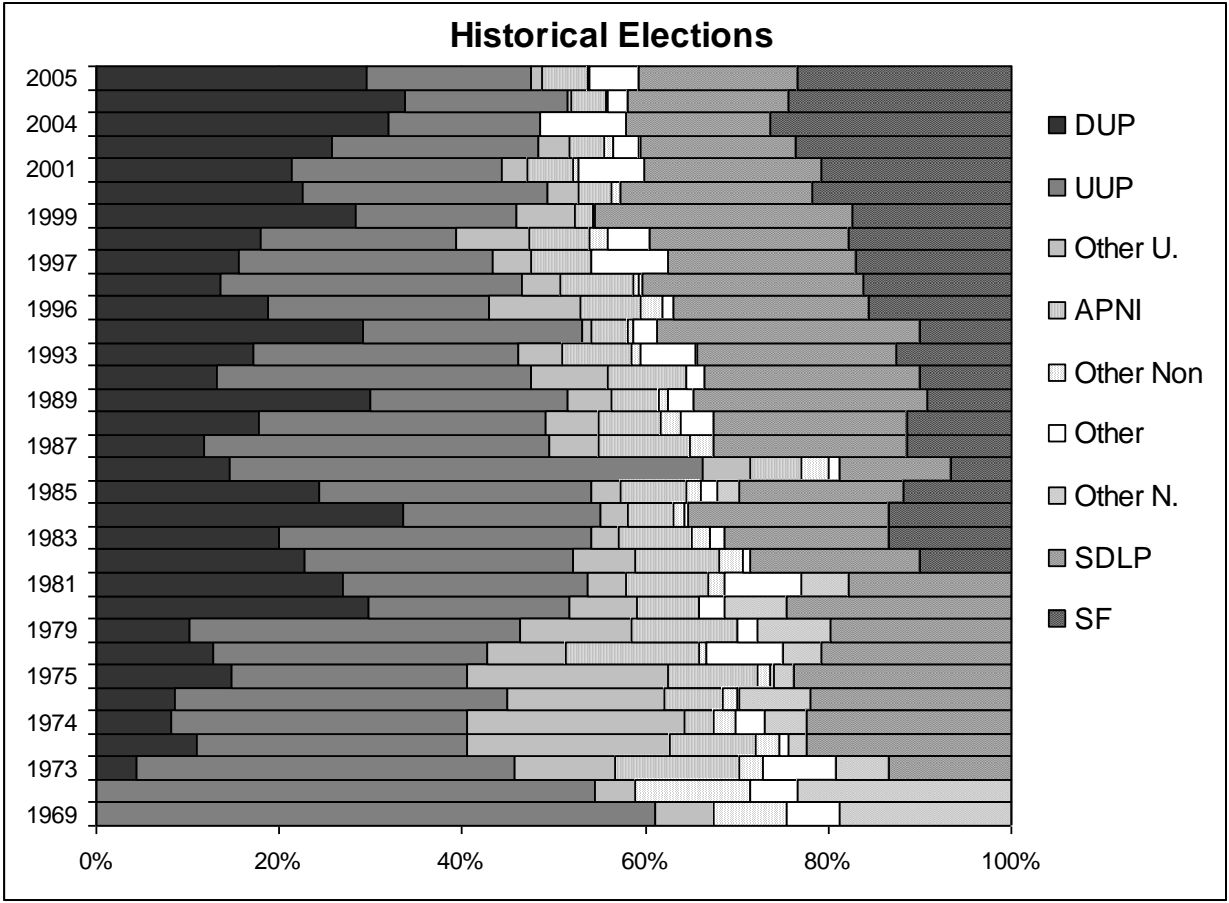

Figure 5. Historical results of elections.

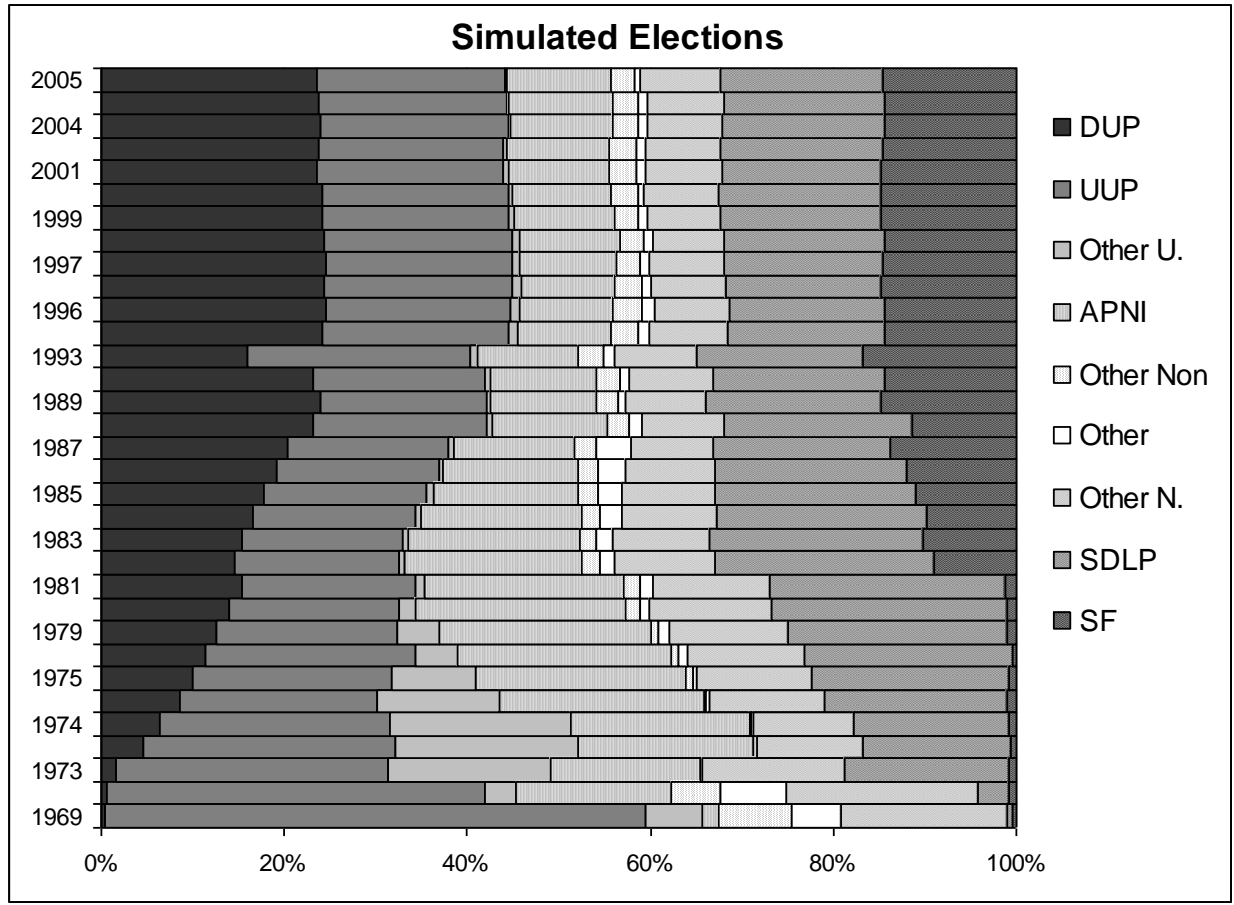

Figure 6. Simulated results of elections.

On the one hand, there is a notable similarity in general trends. For example, the simulated growth trends in the levels of support for Sin Fein post-1981 and for UUP in years 1973-1993 are broadly consistent with actual historical data. The relatively constant levels of simulated support for DUP and SDL are also comparable to real history.

On the other hand, the simulation clearly arrives to a steady state after1994 and seems unable to project any further changes. In our experience, this is a general limitation of the Bound Confident model which tends to converge to an artificial steady state after a period of simulation.

The simulation also strongly overestimates the support to moderate parties, those between UUP and SDLP. Still, it is encouraging to see that an admittedly simple model shows the ability to predict trends in population's sympathies nearly 15 years forward! It is even more remarkable considering that the modelers used very small amount of readily available information: elections results and basic demographics.

As of this writing, extended models of this type, e.g., (Kott and Corpac, 2007; Kott et al, 2007; Kott et al, 2010) are used to study practical problems of modern insurgencies.

## Practical Tips

- Insurgencies are diverse, and no ready-made model will represent all features of the particular insurgency and situation that interest your customer (here, customer is the organization that wishes to use the results of your analysis for applied decision-making). When selecting a model for your analysis, look for one that allows you to make major extensions and modifications.

- A number of simple yet insightful models of insurgency have been published by academic researchers such as the Center for Contemporary Conflict of the Naval Postgraduate School (http://www.nps.edu/Academics/centers/ccc/). Consider using one of such models as an initial baseline, then experiment and gradually extend it for your purposes.

- The ability to trace the chain of influences in analyzing results of a simulation is critically important. System Dynamics models tend to be better in this respect.

- A modeler with little experience in insurgency modeling will find System Dynamics models easier to construct, to understand, and to debug.

- Many insurgencies, and many strategies for managing insurgencies, cannot be properly modeled without representing multiple players—individuals and groups—in the insurgency. In such cases, an agent-based model is most appropriate.

- Always model international actors' influences on insurgents and counterinsurgents. Few insurgencies ever unfold without a major impact by one or more international participant.

- Before constructing a model, ask your analysis's customer about the actions her organization considers taking with respect to the insurgency, or within an insurgency-plagued region. Make sure your model is capable of representing those actions and their effects.

- For every action your customer plans to take, include modeling of undesirable side-effects. Thus, if the analysis' customer plans to provide food aid to refugees, include in your model a possible diversion of food by insurgents.

- In selecting variables and attributes for your model, give preference to most tangible and measurable ones. Include the metrics that your customer expects to use or to affect. Also, consider including historical or current data available to you for calibration and validation.

- Allocate adequate time and resources to review and improve your model with Subject Matter Experts (SMEs). Produce model's visualizations specifically designed for easy comprehension by SMEs.

- Insurgencies are emotional topics, and SMEs tend to hold strong, passionate opinions. Collaborate with several SMEs and welcome widely divergent views. Try to find SMEs who had experiences on both sides of an insurgency.

- Early in the modeling project, work with insurgency SMEs to create a set of test cases. Ask several SMEs, independently, to produce their estimates of probable evolution of the insurgency in each case. Expect to receive widely divergent estimates.

## Resources

Conflict Analysis Resource Center
http://www.cerac.org.co/datasets.htm
Pointers to multiple depositories of datasets related to armed conflicts, inlcuding insurgencies

Ethnicity, Insurgency and Civil War Project
http://www.stanford.edu/group/ethnic/publicdata/publicdata.html
data relating ethnic fractionalization and insurgency

The American Political Science Association, Task Force on Political Violence and Terrorism
http://www.apsanet.org/content_29436.cfm
Pointers to multiple depositories of datasets related to political violence, inlcuding insurgencies

David C. Gompert and John Gordon IV, "War by Other means," RAND Publication 2008
http://www.rand.org/pubs/monographs/MG595.2/
Appendices A and B provide data and analysis of outcomes and correlations for 89 insurgencies, and data on counterinsurgency capabilities of world states and organizations

Political Instability Task Force
http://globalpolicy.gmu.edu/pitf/pitfdata.htm
http://globalpolicy.gmu.edu/pitf/pitfpset.htm
Multiple datasets related to internal wars, including insurgencies

Minorities at Risk Data
http://www.cidcm.umd.edu/mar/data.asp
Datasets characterizing multiple minorities, their conditions and potential risks, including potential or ongoing insurgency

Genocide and Politicide project
http://globalpolicy.gmu.edu/genocide/
Data describing cases of genocide, many related to insurgencies

War and Health website
http://warandhealth.com/civilian-victims-in-an-asymmetrical-conflict-data/
Civilian victims in an asymmetrical conflict

http://www.systemicpeace.org/inscr/inscr.htm
Armed Conflict and Intervention (ACI) Datasets

http://www.systemicpeace.org/
multiple datasets on state instability and conflict

Correlates of War Project (COW)
http://www.umich.edu/~cowproj/dataset.html

Wolrd Bank Datasets
http://econ.worldbank.org/WBSITE/EXTERNAL/EXTDEC/EXTRESEARCH/EXTPROGRAMS/EXTCONFLICT/0,,contentMDK:20336174~menuPK:637270~pagePK:64168182~piPK:64168060~theSitePK:477960,00.html
Data on civil wars post-WWII

Center for Computational Analysis of Social and Organizational Systems (CASOS) at
Carnegie Mellon University. URL:
http://www.casos.cs.cmu.edu/computational_tools/tools.html
Listings and/ or repositories of software tools and libraries

International Network for Social Network Analysis (INSNA). URL:
http://www.insna.org/software/index.html
Listings and/ or repositories of software tools and libraries

Network Workbench (NWB) Tool
https://nwb.slis.indiana.edu/community/?n=Main.NWBTool
network analysis, modeling, and visualization toolkit

# References

Andrew Silke, Rebel's dilemma: The changing relationship between the IRA, Sinn Féin and paramilitary vigilantism in Northern Ireland, Terrorism and Political Violence, Volume 11, Issue 1, 1999, Pages 55 – 93

Jonathan Tonge, Northern Ireland, Polity 2006

Rainer Hegselmann, Ulrich Krause, OPINION DYNAMICS AND BOUNDED CONFIDENCE MODELS, ANALYSIS, AND SIMULATION, Journal of Artifical Societies and Social Simulation (JASSS) vol.5, no. 3, 2002
http://jasss.soc.surrey.ac.uk/5/3/2.html

Friedkin, N. E. (1999). Choice Shift and Group Polarization. *American Sociological Review, 64*(6), 856-875.

Elbadawi, I. and Sambanis, N., "External Interventions and the Duration of Civil Wars," Working Paper 2433, The World Bank, September 2000

D. DUONG, R. MARLING, L. MURPHY, J. JOHNSON, M. OTTENBERG, B SHELDON, S STEPHENS, "NEXUS: AN INTELLIGENT AGENT MODEL OF SUPPORT BETWEEN SOCIAL GROUPS," in Proceedings of the Agent 2007 Conference on Complex Interaction and Social Emergence, Evanston, Illinois, November 15-17, 2007, pp.241-246 isee systems, Stella Software Website, 2009, http://www.iseesystems.com/softwares/Education/StellaSoftware.aspx

MICHAEL DZIEDZIC, BARBARA SOTIRIN, JOHN AGOGLIA, MEASURING PROGRESS IN CONFLICT ENVIRONMENTS (MPICE), Report ADA488249, US Army of Corps of Engineers, 2008

Gabriel Marcella, THE UNITED STATES AND COLOMBIA:THE JOURNEY FROM AMBIGUITY TO STRATEGIC CLARITY, Strategic Studies Institute of the U.S. Army War College, May 2003, http://www.strategicstudiesinstitute.army.mil/pdffiles/PUB10.pdf

P. W. Singer, Children at War, University of California Press, 2006, p. 46

Parkman, J., (2005). Peace Support Operations Study. McLean, Virginia, MORS Workshop on Agent-Based Models and Othyer Analytical Tools in Support of Stability Operations.

Hartley 2008 http://www.mors.org/meetings/es_2008/pres/hartley_dime.pdf

John Benedict, L. Dean Simmons, Enterprise-Wide Opportunities for, Advancing Irregular Warfare Analyses, 13 December 2007 Brief at MORS Workshop

Collier, P., Hoeffler, A. and Söderbom, M. 2001. "On the Duration of Civil War." Policy Working Paper 2861, World Bank, Washington DC.

Mark Abdollahian, Michael Baranick, Brian Efird, and Jacek Kugler, A Predictive Political Simulation Model, Center for Technology and National Security Policy, National Defense University, 2005 http://www.ndu.edu/ctnsp/Def_Tech/DTP%2032%20Senturion.pdf

Sentia Group, Implications of a U.S. Drawdown in Iraq, July 2008, Report, Sentia Group, Inc., http://www.sentiagroup.com/pdf/SentiaInsightMonthly-USDrawdownInIraq-July2008.pdf

Sterman, John D. (2001). "System dynamics modeling: Tools for learning in a complex world". *California management review* **43** (1): 8–25.

Jason Campbell, Michael O'Hanlon, Jeremy Shapiro, Assessing Counterinsurgency and Stabilization Missions, POLICY PAPER Number 14, May 2009, The Brookings Institution Washington, D.C.

JAMES CLANCY and CHUCK CROSSETT, Measuring Effectiveness in IrregularWarfare, Parameters, Summer 2007, pp.88-100

Jason H. Campbell and Jeremy Shapiro, Afghanistan Index: Tracking Variables of Reconstruction and Security in Post-9/11 Afghanistan, a website, http://www.brookings.edu/about/programs/foreign-policy/afghanistan-index

Michael E. O'Hanlon, Jason H. Campbell, Iraq Index: Tracking Variables of Reconstruction & Security in Post-Saddam Iraq, a website, http://www.brookings.edu/iraqindex

F. C. Cuny and R. B. Hill, "Famine, Conflict and Response: a Basic Guide," Kumarian Press, West Hartford, Connecticut, 1999

Nagl, John A. Learning to Eat Soup with a Knife: Counterinsurgency Lessons from Malaya and Vietnam. Chicago: University of Chicago Press, 2002.

Lawrence, T. E. "Evolution of a Revolt." *Army Quarterly and Defense Journal*. Tavistock, United Kingdom (October 1920). Accessed via Command and General Staff College, US Army, Ft Leavenworth KS, at http://cgsc.leavenworth.army.mil/carl/resources/csi/ Lawrence/lawrence.asp on 23 April 2008.

Lawrence, T. E. *Seven Pillars of Wisdom: A Triumph.* Oxford: Alden Press, 1935.

Leites, Nathan and Wolf, Charles. *Rebellion and Authority: An Analytical Essay on Insurgent Conflict*. Arlington: RAND Corporation, 1970 (R-462-ARPA).

Long, Austin. *On "Other War" Lessons from Five Decades of RAND Counterinsurgency Research*. Arlington: RAND Corporation, 2006.

Galula, David. *Counterinsurgency Warfare: Theory and Practice*. Westport, CT: Praeger, 1964.

Peters, Ralph. "The hearts-and-minds myth - Sorry, but winning means killing." *Armed Forces Journal* (September 2006). Accessed via News Bank, Inc. - Armed Services and Government News at http://infoweb.newsbank.com/iw-search/we/InfoWeb on 7 March 2008.

Pierson, Brett M. "OSD/Joint Special Session: Irregular Warfare Activities in OSD and the Joint Staff," *Proceedings of the 76th Annual Military Operations Research Society Symposium.* Alexandria, VA: MORS, 2008.

Edward G. Anderson Jr,  A Proof-of-Concept Model for Evaluating Insurgency Management Policies Using the System Dynamics Methodology Strategic Insights, Volume VI, Issue 5 (August 2007)

Baker, A., "How the Taliban Thrives," Time, vol.174, no.9, September 7, 2009, pp.46-53

Choucri, N., C. Electris,  D. Goldsmith, D. Mistree, S. Madnick, J. Morrison, M. Siegel,  & M. Sweitzer-Hamilton. 2006. "Understanding & Modeling State Stability: Exploiting System Dynamics." Proc. of 2006 Institute of Electrical and Electronics Engineers Aerospace Conference. Big Sky, MT: IEEE, 2006.

US Army, Counterinsurgency Field Manual FM-3-24, Paladin Press, 2006

Hegre, H., T. Ellingsen, S. Gates, N. Gleditsch, "Toward a Democratic Civil Peace? Democracy, Political Change, and Civil War, 1816-1992," *American Political Science Review*, Vol. 95, Issue 1, March, 2001: 33-48. Available online: http://www.worldbank.org/research/conflict/papers/CivilPeace2.pdf

Iyengar, R., & Monten, J. (2008). *Is there an" emboldenment" effect? Evidence from the insurgency in Iraq* (No. w13839). National Bureau of Economic Research.

JAMES T. QUINLIVAN, Force Requirements in Stability Operations, From Parameters, Winter 1995, pp. 59-69.

Alex Grynkewich, USAF and Chris Reifel, USAF, Modeling Jihad: A System Dynamics Model of the Salafist Group for Preaching and Combat Financial Subsystem, Strategic Insights, Volume V, Issue 8 (November 2006)

Tara A. Leweling, USAF and Otto Sieber, USN, Calibrating a Field-level, Systems Dynamics Model of Terrorism's Human Capital Subsystem: GSPC as Case Study, Strategic Insights, Volume V, Issue 8 (November 2006)

Michael Kowalchuck, Siddhartha Singh and Kathleen M. Carley, CORES – Complex Organizational Reasoning System, Report CMU-ISRI-04-131, Carnegie Mellon University, School of Computer Science, September 2004

Robbins, M. (2005). *Investigating the Complexities of Nationbuilding: A Sub-National Regional Perspective.* Master Thesis, AIR FORCE INSTITUTE OF TECHNOLOGY, Wright-Patterson Air Force Base, Ohio

US Department of Defense (12 July 2007) (PDF), Joint Publication 1-02 Department of Defense Dictionary of Military and Associated Terms, JP 1-02, http://www.dtic.mil/doctrine/jel/new_pubs/jp1_02.pdf
 retrieved 2007-11-21

Hegselmann, R., & Krause, U. (2002). Opinion Dynamics and Bounded Confidence Models, Analysis, and Simulation. *Journal of Artificial Societies and Social Simulation, 5*(3).

Friedkin, N. E. (1999). Choice Shift and Group Polarization. *American Sociological Review, 64*(6), 856-875.

CAIN Web Service (May, 2006). *Political party support in Northern Ireland, 1969 to the present*. http://cain.ulst.ac.uk/issues/politics/election/electsum.htm

Grier, R. A., Skarin, B., Lubyansky, A., & Wolpert, L. (2008). SCIPR: A computational model to simulate cultural identities for predicting reactions to events. Proceedings of the Second International Conference on Computational Cultural Dynamics

Kott, A., & Corpac, P. S. (2007). *COMPOEX technology to assist leaders in planning and executing campaigns in complex operational environments*. 12th International Command and Control Research and Technology Symposium, 2007

Kott, A., Hansberger, J., Waltz, E., & Corpac, P. (2010). *Whole-of-Government Planning and Wargaming of Complex International Operations: Experimental Evaluation of Methods and Tools*. International Journal of Command and Control, The International C2 Journal, Vol 4, No 3, March 2010

Kott, A., Hawley, L., Brown, G., Citrenbaum, G., & Corpac, P. S. (2007). *Next State Planning: A" Whole of Government" Approach for Planning and Executing Operational Campaigns*. 12th International Command and Control Research and Technology Symposium, 2007

Kott, A., & Citrenbaum, G. (Eds.). (2010). *Estimating Impact: A Handbook of Computational Methods and Models for Anticipating Economic, Social, Political and Security Effects in International Interventions*. Springer Science & Business Media.